# 17. The Well-tempered Compiler?
# The Aesthetics of Program Auralization

Paul Vickers & James L. Alty

## 17.1.    Introduction

> *Like angels stopped upon the wing by sound*
>
> *Of harmony from Heaven's remotest spheres.*
>
> —Wordsworth: The Prelude

In this chapter we are concerned with external auditory representations of programs, also known as program auralization. As program auralization systems tend to use musical representations they are necessarily affected by artistic and aesthetic considerations. Therefore, it is instructive to explore program auralization in the light of aesthetic computing principles. In The Music of The Spheres, James (1993) writes of music and science that "*at the beginning of Western civilisation … the two were identified so profoundly that anyone who suggested that there was any essential difference between them would have been considered an ignoramus.*" This is in stark contrast to today when anyone suggesting that they have anything in common "*…runs the risk of being labelled a philistine by one group and a dilettante by the other and, most damning of all, a popularizer by both.*"

The Great Theme beloved of the early philosopher scientists, of a universe of perfect order in which everything has a purpose and a place, a universe whose very fabric sounded to continual heavenly music (which music obeyed the beautiful rules of the mathematics of Pythagoras and Plato), was discarded over the years of the Renaissance and into the Age of Reason. Though many present-day scientists have a great appreciation of the arts, those involved in the humanities often eschew the cold empiricism of science. This is the age of C.P. Snow's Two Cultures[1], which James describes as a "*psychotic bifurcation*". James elaborates:

> In the modern age it is a basic assumption that music appeals directly to the soul and bypasses the brain altogether, while science operates in just the reverse fashion, confining itself to the realm of pure ratiocination and having no contact at all with the soul. Another way of stating this duality is to marshal on the side of music Oscar Wilde's dictum that 'All art is quite useless,' while postulating that science is the apotheosis of earthly usefulness, having no connection with anything that is not tangibly of this world.

Despite centuries of divergence, there are indications that some are starting to build bridges between the cultures again. In Douglas Adams' comedy novel Dirk

Gently's Holistic Detective Agency (1988), the lead character Richard MacDuff attempts to produce music from mathematical representations of the dynamics of swallows in flight. In a marvellous reiteration of the Great Theme, Adams writes of MacDuff's belief that "*if ... the rhythms and harmonies of music which he found most satisfying could be found in, or at least derived from, the rhythms and harmonies of naturally occurring phenomena, then satisfying forms of modality and intonation should emerge naturally as well*."

Although a single work of comic fiction is not scientific evidence of a swing away from The Two Cultures, Adams' thinking is indicative of a growing trend in the computer science research community. It is doubtful that MacDuff's beliefs were an intentional move by Adams towards re-establishing the philosophy of the early scientists among modern researchers; however, the fact remains that computer scientists (unwittingly or not) are making increasing use of artistic forms (be they aural or visual) in their work.

## 17.2.    Auditory display

Since the introduction of the visual display unit much research effort has gone into finding new and better ways to maximise the use of the video channel. Developers have been quick to maximise the use of graphical display capabilities from the use of menus on character-based displays to the visually impressive graphical user interfaces of today.

Alongside the development of visual presentation, psychologists spent much time analysing and studying the effects on computer users of different methods of information display. This has led to a well-established body of research into the exploitation of the visual medium as a means of interfacing the increasingly powerful and sophisticated computer technology with a more discerning and expectant user community.

Although in the early days of computing simple audio signals were experimented with, the research community was slow to recognise audio as a useful carrier of information in the world of software development. This can partially be attributed to the relatively late arrival of widely available and cheap sound generating devices for computers. By the time affordable sound generating equipment became available to the average computer user the study of the visual medium was well advanced. For largely technological reasons the human-computer interface has from the start been almost entirely visual in its construction. With the advances in display technology came an inertia that led to an increasing bias towards visual interfaces. This is reflected in the natural language of those cultures that rely on the written word for communication (particularly English), which, by using words like 'imagery' to describe mental processes, shows an inclination towards visual metaphors for the explanation of ideas. Contrast this with cultures that have an oral tradition which are much more multi-sensory in their communication:

Speakers in non-literate cultures, including children in all cultures who have not yet learned to read, tend to use many inflections and gestures. But as people become educated in literate cultures they are often taught to "modulate" their vocal inflections, stand still as they talk, and not use gestures. Thus speech becomes reduced to the single element which can be coded by writing or printing: the meaning of the words themselves (Somers, 1998).

In Dirk Gently's Holistic Detective Agency, MacDuff made himself wealthy by devising a spreadsheet program that allowed company accounts to be represented musically. MacDuff states sarcastically that the "*yearly accounts of most British companies emerged sounding like the Dead March from Saul*" (Adams, 1988). Although Adams' idea of the musical spreadsheet may have seemed absurd in 1988, science fiction often precedes science fact and researchers such as Kramer (1994a) have since reported the successful use of auditory displays of stock market data to identify market trends.

The idea of auditory imagery has begun relatively recently to attract attention in the cognitive fields (e.g. see Reisberg, 1992). Where graphical visualisation is informed by complementary research in cognitive science, auditory display draws upon corresponding auditory image research in addition to the audio engineering/sound production fields to allow the communication of information and data through non-speech sound (see Figure 17.1). Auditory imagery is, of course, vital for blind users. Up until the 1970s blind people were quite involved in computing. The selective development of visual interfaces was a severe blow to them.

< Insert Figure 17.1>

Although not always explicit in the literature, auditory display work is also informed to a greater or lesser extent by aesthetic considerations.

## 17.3. The programming problem

Computer programming poses an interesting problem for information display. Program events are in the time domain whilst visual mappings provide predominantly spatial representations. Visual techniques give us good descriptions of spatial relations and structural details (just like Fourier analysis does for sound waves), but do not naturally represent temporal details. Sound presents us with a complementary modality that increases the diagnostic tools available by giving a temporal view of software (as the wave-form plot does for a sound wave). In fact, audio was used quite a lot in the early days of computing. Machines such as the ICL 1900 series had sound output on their operator consoles and it was often used by operators and engineers who, by listening to the patterns of sounds from the loudspeaker, learnt to monitor CPU behaviour and to identify errant program behaviour. However, most sound output was primitive and required much effort to produce.

Jackson and Francioni (1992) argued that some types of programming error (such as those that can be spotted through pattern recognition) are more intuitively obvious to our ears than our eyes. Also, they pointed out that, unlike images, sound can be processed by the brain passively, that is, we can be aware of sounds without needing to actively listen to them. The representation of program information in sound is known as program auralization (Kramer, 1994b). One of the first auralization systems was described by Sonnenwald et al (1990) and was followed by DiGiano (1992), DiGiano and Baecker (1992), Brown and Hershberger (1992), Jameson (1994a; 1994b), Bock (1994; 1995a; 1995b) Mathur et al (1994), and Boardman et al (1995). These early systems all used complex tones in their auditory mappings but, like much other auditory display work, this was done without regard to the *musicality* of the representations. That is, simple mappings were often employed, such as quantising the value of a data item to a chromatic pitch in the 128-tone range offered by MIDI-compatible tone generators. Furthermore, the pitches were typically atonal in their organisation and were combined with sound effects (e.g. a machine sound to represent a function processing some data). Effort was largely invested in demonstrating that data could be mapped to sound with much less attention given to the aesthetic qualities of the auditory displays. Alty (1995) was one of the first to explicitly use musical principles in his auralization of the bubble-sort algorithm. Leyton (see chapter 15 of this volume) argues that strong aesthetics maximize the transfer of structure. Indeed, where aesthetic considerations are taken into account auditory displays become much easier to listen to and to comprehend (as evidenced by Alty, 1995) as the transfer of information from the computer domain to the auditory domain is facilitated. Mayer-Kress et al (1994) mapped chaotic attractor functions to musical structures in which the functions' similar but never-the-same regions could be clearly heard. The use of a musical aesthetic meant that the resultant music could be appreciated in its own right without needing to know its generative history (how it was produced).

## 17.4.    Music in auralizations

In his *Seismic Sonata* Quinn (2000) used the aesthetics of tonal musical form to sonify data from the 1994 Northridge, California earthquake. Using data to assist with composition is not new. Cohen (1994) suggested that it was John Cage who first put forth the principles of auditory display in the 1950s, citing Cage's works *Music of Changes* (1952) and *Reunion* (1968) as early examples of data sonification. In Music of Changes, the score was written by mapping the results of coin tosses to pitch, duration, amplitude, and timbre. Even changes in tempo and the number of measures in a given section were controlled by coin tosses. Reunion developed the idea by using photoelectric switches on a chessboard to trigger the playing of different pieces of music. Whether Cage intended to communicate information regarding data sets by music or whether he merely used data as a mechanism for the creation of new music (i.e., was the music a by-product or the intentional product) is a moot point; what is interesting is that Cage believed the relationship between music and data could be exploited. Indeed, King and Angus (1996) believed that musical aesthetics would provide a sufficiently well-understood framework upon which to build a useful auditory display of the DNA gene sequence of the brain's serotonin receptors that the resultant sonification also appeared as the CD album track *S2 Translation* (The Shamen, 1995).

Vickers and Alty (2002b) argued that music offers a powerful medium for communication and so looked for ways to use its structures and organisational principles to better communicate program information. Francioni et al (1991) suggested that musical representation can highlight situations that could easily be missed in a visual representation (and no doubt there are also cases where the opposite is true). To give a simple example, shifting a single note by one semitone can change the whole sense of a chord and produce an immediate and compelling effect. This happens when a major triad has its mediant (the third degree of the scale) flattened to produce a minor chord (e.g. see Figure 17.2): a similar movement in the value of one data variable in a graph might not be noticed. Of course, not all semi-tone shifts would be as dramatic as they may merely serve to colour a chord rather than change its type. However, the fact remains that, within a tonal music framework, very small changes in pitch can be readily discernable when they change the balance of the melody (or the melodic contour) and can even sound out of place by falling outside the organizing rules of the particular musical style. Thus, perturbations in the data being explored can be mapped to musical events that are easily perceived, and so the aesthetics of tonal music increase the transferability of this information to the listener.

< Insert Figure 17.2>

Figure 17.2 (a) shows a C-major triad in first inversion form in which the bottom note is the mediant E. In (b), the E is flattened changing the chord into a C-minor triad (also in first inversion form). Although the change is small (a frequency shift of approximately 6% from 329.6Hz to 311.1Hz) the effect is very noticeable. In program debugging the richness provided by a musical representation may offer fairly precise bug location possibilities (whether used in isolation or in conjunction with the visual media).

The key issue is how to map domain entities to musical structures. Alty (1995) showed that algorithms (such as the bubble sort and minimum-path) can have information about their run-time behaviour communicated successfully through musical mappings. The results suggest that, provided precise numerical relationships are not being communicated, music can transfer information successfully. In their development of a musical diagram reader for the visually impaired, Alty and Rigas (1998) concluded that musical messages should be designed within a consistent framework (much as elements of successful graphical user interfaces follow common design principles). With the CAITLIN musical program auralization system Vickers and Alty (2002a; 2002b; 2002c; 2003) demonstrated that a musical auralization framework for communicating run-time behaviour of Pascal programs was successful in assisting with bug location tasks.

In the CAITLIN system, motifs (signature tunes) were composed for the program language features to be displayed (in this case the language constructs WHILE, REPEAT, FOR…TO, FOR…DOWNTO, IF, IF…ELSE, CASE, and CASE…ELSE). The motifs were organised around a unified and structured tonal framework (see Vickers & Alty, 2002a) and their design was strongly influenced by current thinking in music theory and music cognition. The aim was to create a musical environment that would be easily and quickly learnt, that was not dependent on prior musical training or expertise, and that could thus be used to communicate information about the run-time behaviour of a program. Diatonic (seven-note scale) forms were used as these underlie much western popular and orchestral music and so already have wide exposure in the general population from which the experimental subjects were drawn. Figure 17.3 shows Pascal code for a loop and a selection construct and the resultant auralizations.

< Insert Figure 17.3>

Until programs become self-aware and can identify their own bugs, it is left to the programmer to diagnose the symptoms of a mal-functioning program and deduce where its defects lie. Thus, the auralizations themselves do not have musical features that represent bugs: a bug causes a perturbation in program flow and it is these perturbations that are looked for. For example, the auralization in Figure 17.3 (d), which was generated from the code in Figure 17.3 (c) shows that the variable 'a' did not have any of the values 1, 2, or 3. If, at this point in the program, 'a' was supposed to have a value in the range 1..3 then either the statement that assigns a value to 'a' is in error, or the earlier statement that gives 'b' a value (not shown) is in error: either way, the fact that 'a' does not have an expected value manifests itself in the auralization which tells the listener that the CASE statement's ELSE path was followed. If 'a' did have a value in the range 1..3 then the auralization would sound different as we would hear a major motif signifying a match, as in Figure 17.3 (e)[2].

## 17.5.    An aesthetic perspective on auralization

With the CAITLIN system we showed that simple diatonic auralizations were useful (Vickers & Alty, 2002c; 2003). One could argue that, given sufficient training, any auralization framework can be learnable and usable. However, that goes against the more recent efforts to improve design aesthetics, as espoused and championed, for example, by Norman (2004). The underlying principle of aesthetic computing is that art theory and practice should influence the design of computer systems and artefacts (Fishwick, 2002). As art theory and practice embody customization, personalization, preference, culture, and emotion (Fishwick, 2003, and chapter 1 of this volume) there is much for the auralization designer to consider.

### 17.5.1.  Customization and personalization

Many of the early program auralization systems let the user define the mappings of data to sound. Whilst this allows almost unlimited customization, personalization, and preference, it requires some sound-design skills on the part of the user. It's like giving someone the twelve notes of the chromatic scale and asking them to compose a sonata — without knowledge of, and training in, composition techniques this would be a near-impossible task. Customization and personalization must be balanced by the knowledge and skills required to make use of them. In the CAITLIN system we took the opposite approach and decided to use fixed auralizations, such as those shown in Figure 17.3. The motifs were pre-assigned to the various language constructs as they had been designed hierarchically so that all selections were variations on a common theme and all iterations were variations on a different common theme. The system allows the timbre for each construct to be altered and in some cases the musical scale (e.g.

major, minor, ten-note blues, etc.) can also be selected (notably for the FOR loops). However, in the experimental setting (Vickers & Alty, 2002a, 2002c), only the overall tempo was user-adjustable so as not to confound the results.

### 17.5.2. Preference

User preference is certainly an important factor. A study of surgeons who listened to music while operating showed that their speed and task accuracy were greater when they listened to self-selected music rather than music chosen by the experimenters (Allen & Blascovich, 1994). In our experiments we noticed a definite preference amongst subjects for motifs with a strong melody. This was especially apparent in the early pilot studies where the motif design was less refined. In the early versions of the system some constructs had motifs that were much less musical than others. For example, the first version of the system used a metaphoric mapping for the IF and IF…ELSE statements: a pitch bend was applied to mimic the rising and falling inflection of the human voice when posing and answering questions. In fact, as several of the subjects in the experiment observed, this ended up sounding like a comical ship's fog horn (Vickers, 1999). In the most recent version users expressed a preference for the FOR motifs which were also the most melodic, that is, the melodic contour was more elaborate than the simpler up-and-down scale-based motifs of the selection constructs and the harmonically-richer motifs of the WHILE and REPEAT loops. Contour was judged by subjects in our studies as being a useful aide-mémoire for recalling the motifs. Edworthy (1985) and Dowling (1982) observed that contour becomes even more important when the tonal context is weak or confusing; contour becomes less important in familiar melodies and melodies retained over a period of time.

### 17.5.3. Emotion

Music clearly has an emotional dimension. We talk of mood music and can be strongly moved by certain pieces. One emotion that auralization systems are susceptible to induce is annoyance. Gaver and Smith (1990) noted that sounds in the interface can be annoying and that what "seems cute and clever at first may grow tiresome after a few exposures". In our experience, tiresome sounds are usually those that have not been designed with listening aesthetics in mind – that is the mental and cognitive processing loads required by the sounds themselves reduce the amount of attention that can be given by the listener to the information transfer function. Designers go to great lengths to ensure that auditory signals and alarms in safety critical environments (such as aircraft cockpits and nuclear power plants) sit well within their auditory ecology, but these rigours are not so well followed in other auditory displays. It is easy to make a display that clashes with, or masks, other events, or that is simply tiring (both emotionally and cognitively) to use. Approximately half of the subjects in an experiment using the CAITLIN system found the auralizations to be moderately annoying, the other half suffering almost no annoyance (Vickers, 1999). The ambiguity of this result gives hope that auralizations can be produced that do not trigger a negative emotional response but cautions us that we must pay very careful attention to this aspect.

### 17.5.4. Cultural aspects

Early auralization systems used musical pitches and MIDI data, but they were simply mapping program data to common frequencies to effect the auralizations without regard to the musicality of the output. Weinberg (1998) described programming as a "communication between two alien species" and Conner and Malmin (1983) said that we must recognise that a gap in understanding may exist between the communicator and the receiver. For successful communication there must be a common medium between the two in order that the gap may be bridged. Meyer (1956) observed that meaning and communication "cannot be separated from the cultural context in which they arise. Apart from the social situation there can be neither meaning nor communication." Music aesthetics are thus culturally dependent and so the aesthetics of an auditory display have a pivotal role in determining how successful the display is. Watkins and Dyson (1985) found that music performed in a style familiar to the listener is easier to recognise and understand. In the CAITLIN system it was vital that the auralizations were not so far from the programmer's frame of reference as to be rendered useless. If music is to be used, it must not rely on forms and intervals that are too unfamiliar or indistinguishable to the average person. That is, the aesthetics must be complementary with, or accessible to, those of the listener. Composers organise music according to defined structures, schemas, or sets of rules. Structuring auralizations according to simple syntactical rules offers the hope of music forming the basis for a bridge of the semantic gap between an incorrectly functioning program and the programmer.

Alty (2002) observed that just as designers would never create a chair twelve metres high because it would not be generally useable, composers must not produce works that are beyond the cognitive processing capabilities of the listener. For example, some composers have chosen to use transformations which are simply not cognitively identifiable. In the same vein, auralizations must be mappable to different musical idioms so that the user can select a representation that is familiar. Just as software interfaces undergo internationalization to take account of cultural differences and social constructs, so auralizations need to be designed with the listener in mind. What is particularly interesting about music is that recall of melody appears to be an innate skill. That is, people do not need to be trained to recognise melodies (the ability to sing, whistle, or hum a tune after only a few hearings is evidence of this). Thus, auralizations that use melodies as carriers of information stand a good chance of being understood and retained in the mind of the listener.

The diatonic scale is so common in western music that one can be fooled into thinking it is somehow a form of nature. But, as Parncutt (1989, p. 5) observed, it is not "*an inevitable consequence of the psychophysics of tone perception*". In the nineteenth century Helmholtz believed that the development of musical styles was heavily influenced by culture and aesthetics. This is evident in the divergence of the eastern and western musical traditions. The western classical tradition (especially in the eighteenth century) was driven by a desire to explore harmony. Eastern music, on the other hand, focused less on harmony and much more on rhythmic structures (see Parncutt, 1989 p. 6). In the ancient Greek world there was great debate about the relative spiritual merits and vices of the different modal schemes (scales) that were common right up until the middle ages.

The argument that diatonic systems are in some way more natural than atonal systems is belied by the fact that concert repertoires continue to include new music styles. However, as Parncutt (1989) observes, most of the atonal systems have not been incorporated into mainstream (or popular) music as they require more information processing by the listener; studies have shown that the organising principles of twelve-tone music in which there is no tonal centre to the music and each degree of the scale has equal weight (for example, as practised by Schoenberg and Stockhausen) are often imperceptible even to trained listeners. Alty (2002) offered an explanation of this in terms of the limits of working memory which, according to Miller (1956), can handle around seven concurrent bits (or chunks) of information. In experiments on melody recall Sloboda and Parker (1985) found that the most fundamental feature preserved in a recalled melody was its metrical structure. Musicians and non-musicians differed significantly only on one measure, that of the ability to retain the harmonic structure of the original melody. Therefore, it is wise not to rely on ability to discriminate between harmonic structures in the auralization motifs, and so we do not commend atonal music systems as good vehicles for auralization.

Of course, there are cultural as well as perceptual factors at work here as the seven-note diatonic tonal scheme is a western, not a world, music form. That said, there is evidence to suggest that the scheme shares characteristics with other world music forms. For instance, melodies from around the world tend to centre on a particular pitch, thus exhibiting a key feature of tonality (Parncutt, 1989, p.70). Furthermore, the twelve-note chromatic scale (of which the diatonic scale is a subset) developed independently in different musical cultures (ancient China, India, Persia, and then the west) and the use of the octave, fourth, and fifth intervals (important in tonal forms) is widespread in world music. Besides, the international success of western rock and pop bands is evidence that even western musical structures are widely (if not universally) accepted, especially in the computer-using world (Vickers & Alty, 2002b).

Nevertheless, as designers of auditory displays we must be aware that even an idiom (such as western pop music) that is widely *accepted* is not necessarily *interpreted* the same way around the world, for the boundary between sensory and cultural influences is not clear. For example, consonance and dissonance are important concepts but ones which appear to be specific to western music (Parncutt, 1989). This means that comprehension (or rather, specific interpretations) of particular musical structures cannot be taken for granted. An auralization system that uses dissonance to draw attention to exceptional events, for example, may fail for listeners who are more influenced by musical forms that do not place the same emphasis on consonance and dissonance. So, we can see that the aesthetic issues of program auralization systems are complex and are strongly culturally dependent.

## 17.6.    Conclusions and Future Work

In an attempt to avoid the pitfalls of requiring programmers to be able to specify good auditory mappings, the CAITLIN musical program auralization system was built with fixed motifs that were designed within a coherent and self-consistent tonal framework (see Vickers & Alty, 2002a). This helped to ensure that the aural ecology of the system was healthy and that no individual parts of the auralizations dominated the mix or conflicted with others. The benefits of this approach are that the listener receives output consistent with a unifying aesthetic framework. A disadvantage is that the system is much less configurable to suit different preferences and emotional or cultural needs. In a sense, an aural equivalent of XML is needed to allow the content (information or data) to be separated from its presentation (in this case, the auditory metaphor). Then designers could produce sets of auditory mappings in much the same way that visual interfaces (or skins) are produced for popular programs today. For example, there could be a jazz schema, a Bach chorale schema, a Javanese Gamelan music set, or even a Chinese classical opera style. In addition, we envisage providing multiple motifs within each set so that program objects and events can be tagged by the user with the motif of preference, in the knowledge that each motif conforms to the aesthetic qualities of the others. Such a development could be considered to be extending the principles of literate programming (Knuth, 1984; Pardoe & Wade, 1988). Where literate programming tools of the past concentrated on typography and external visual representations to enhance presentation and comprehension of programs (e.g. Vickers, Pardoe, & Wade, 1991a, 1991b), the tools of the future can make use of auditory and musical aesthetics to extend the programmer's toolbox and visualization set.

Of course, more experimentation is needed to explore just how sensitive auralizations are to the cultural and aesthetic background of the listener. So far the CAITLIN system has only been tested within a western tonal system with western subjects. Actually, we would be surprised if the simple musical forms of the CAITLIN system were not comprehensible to people from other cultures given the tendency of other world music systems to have tonal characteristics. Indeed, we have found in many other tests with subjects from many countries that the cultural differences are minor if simple forms and structures are used.

In the pursuit of aesthetic excellence we must be careful not to tip the balance too far in favour of artistic form. Much current art music would, perhaps, not be appropriate for a generally-usable auralization system. The vernacular is popular music the aesthetics of which are often far removed from the ideals of the music theorists and experimentalists. Lucas (1994) showed that the recognition accuracy of an auditory display was increased when users were made aware of the display's musical design principles. Watkins and Dyson (1985) demonstrated that melodies that follow the rules of western tonal music are easier to learn, organise (cognitively), and discriminate between than control tone sequences of similar complexity. So, it would seem that the cognitive organisational overhead associated with the aesthetics of atonal systems makes them less well suited as carriers of program information.

The sonifications of Quinn (2000) and King and Angus (1996) and the generative music of Mayer-Kress et al (1994) had a dual function to stand on their own as music but also to shed light on the underlying data. In a sense, a program auralization system is a generative music system in that the musical output is dependent on the input data; changing the data changes the behaviour of the program and thus the music. However, the purpose of auralization systems is not to entertain or to convey mood and emotion, but to assist programmers with understanding software and its behaviour. The intentional product of an auralization system is the communication of information or knowledge with the music as the carrier. The music itself, inasmuch as it exists as an entity in its own right, is not the intentional product but a by-product of the auralization process. Therefore, whatever music systems and aesthetics are employed they must not detract from the prime purpose which is to communicate information. Of course, if the mappings can be organised such that the music by-product can exist and be appreciated independent of its context (as is the case with Mayer-Kress et al's generative chaotic attractor music, for example) then so much the better.

Very few formal studies of program auralization have been published. The experiments described by Vickers and Alty (2002a; 2002c) indicate that music can be used to communicate information about program flow and assist with bug location. The results highlighted two areas where the music seemed particularly efficacious: where the program output contained no clue as to the bugs' location and where programs contained complex Boolean expressions. When the output gives clues (e.g. a loop only displays six output records instead of an expected ten) then a bug's location is relatively easy to guess at because the auralization very quickly showed the loop to be at fault. However, when no such clue exists the job is harder to do. In the case of multiple complex Boolean expressions, auralization made it very easy to hear which expressions were at fault; without the auralization subjects had to evaluate the expressions by hand (or use a visualization).

Auralization support for object-oriented and multi-threaded programming environments is necessary. The potential for musical sound in program comprehension in such a domain needs to be explored. For example, the orchestral model of families of timbres (e.g. woodwind, brass, strings, percussion, and keyboards) could be applied to help programmers to distinguish between the activities of different threads. Furthermore, rather than replacing visual displays (though this would be useful for the visually-impaired) we anticipate that combination audio-visual displays will be the most powerful. We expect that the temporal-spatial communication space provided by a combination auralization-visualization system will offer the programmer a powerful set of tools for writing, comprehending, and debugging code. For example, we envisage a scenario in which a graphical visualization displays the state of a data structure whilst an auralization renders the program's control flow (or, perhaps, the passing of messages between object methods and program threads). Integrating auralization tools into software development environments is necessary to allow common debugging techniques (such as breakpoints and step-and-trace facilities) to be extended into the auditory domain.

The use of a bi-modal system offers exciting opportunities for program comprehension and debugging tasks. The ease with which music and non-speech audio can now be incorporated into programming environments (especially the Java platform) means that such a system is a realisable goal in the short to medium term. As long as the auditory aesthetics are well-designed and thus support the transfer of information from the symbolic programming domain to the temporal auditory domain, we believe such a tool will be a valuable addition to the software development community.

**FIGURES**

FIGURE 17.1: The emerging discipline of auditory display

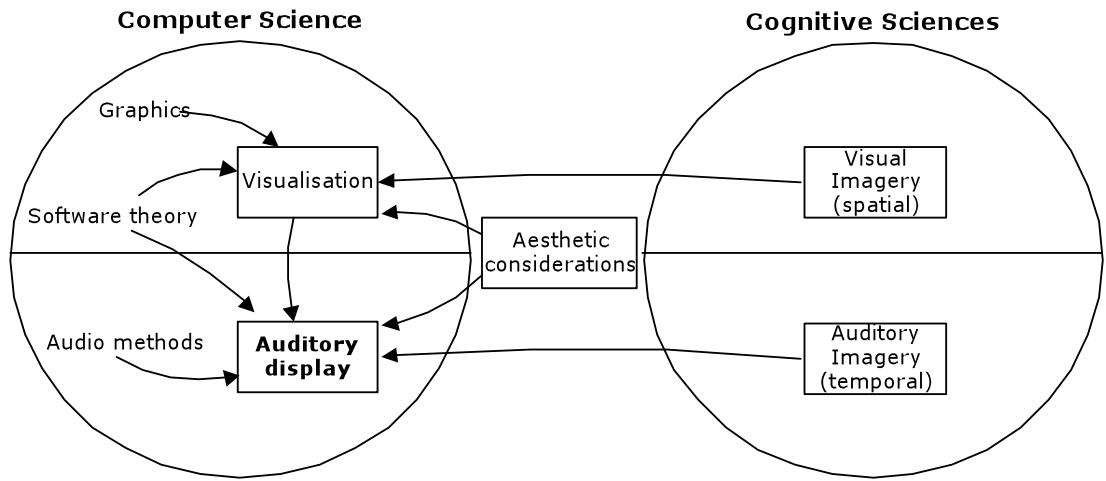

FIGURE 17.2 A semitone shift produces a very noticeable effect. (a) shows a C-Major chord in first inversion (the mediant at the bottom) and (b) shows a C-Minor chord in first inversion.

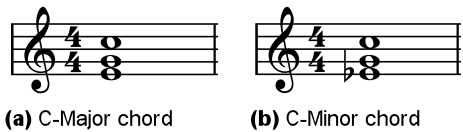

**(a)** C-Major chord          **(b)** C-Minor chord

FIGURE 17.3 CAITLIN motifs for a FOR…TO loop (b) and a CASE…ELSE

statement. Two auralizations for the CASE…ELSE are given: one with a match

(d) and one with no match (e).

**(a)**

```
FOR counter := 1 TO 6 DO
    counter := counter + 1 ;
```

The Pascal code (a) results in the auralisation (b). Bar 1 has a percussive motif played on an open triangle that prefixes all iterations Bar 2 and beat 1 of bar 3 show the tune denoting entry to the loop. The notes in the remainder of bar 3 and in bar 4 represent six iterations of the loop, the final iteration being supplemented by a sleighbell sound. Bars 5 and 6 denote exit from the loop (the closed triangle motif in bar 6 terminates all iterations).

**(b)**

**(c)**

```
a := b + 3 ;
CASE a OF
    '1' : Writeln ('Found 1') ;
    '2' : Writeln ('Found 2') ;
    '3' : Writeln ('Found 3') ;
    ELSE  Writeln ('Not found)
END ;
```

The Pascal code (c) results in the auralisation (d) when the assignment to variable 'a' gives a value greater than 3 and in the auralisation (e) when the assignment gives 'a' the value 3. Bar 1 contains a percussive motif that is prefixed to all selections. Bar 2 and beat 1 of bar 3 show the tune denoting entry to the CASE construct.

**(d)**

A cowbell sound is played as each case instance is tested (bottom staff, bars 3 and 4). The tune in bars 5 and 6 signifies exit from the construct. In this example, no match was found for the case selector (meaning the variable 'a' did not have a value in the range 1...3), so a minor chord (beat 1, bar 5) is played to signal following of the ELSE patch and then the construct finishes with the exit motif in a minor key. In the CAITLIN system the major mode was used to denote Boolean true and minor for Boolean false.

**(e)**

In the auralisation below we can see a major chord on beat 2 of bar 4 played above the cowbell sound that denotes testing the third CASE instance. This means that the value of 'a' matched that of the third CASE instance (i.e., it had the value 3). The construct now exits in a major key (bars 5 and 6) because of a successful match.